\documentclass[fleqn,10pt]{wlscirep}
\usepackage[utf8]{inputenc}
\usepackage[T1]{fontenc}
\usepackage{lineno}
\usepackage{adjustbox}
\usepackage{multirow}

\usepackage{amssymb}
\usepackage{pifont}
\newcommand{\cmark}{\ding{51}}%
\usepackage{subfigure}
\def\CTMD{\text{COVID-CT-MD}}
\usepackage[edges]{forest}

\definecolor{folderbg}{RGB}{124,166,198}
\definecolor{folderborder}{RGB}{110,144,169}

\def\Size{4pt}
\tikzset{
  folder/.pic={
    \filldraw[draw=folderborder,top color=folderbg!50,bottom color=folderbg]
      (-1.05*\Size,0.2\Size+5pt) rectangle ++(.75*\Size,-0.2\Size-5pt);
    \filldraw[draw=folderborder,top color=folderbg!50,bottom color=folderbg]
      (-1.15*\Size,-\Size) rectangle (1.15*\Size,\Size);
  }
}

\title{COVID-CT-MD: COVID-19 Computed Tomography (CT) Scan Dataset Applicable in Machine Learning and Deep Learning}

\author[1]{Parnian Afshar}
\author[2]{Shahin Heidarian}
\author[1]{Nastaran Enshaei}
\author[1]{Farnoosh Naderkhani}
\author[3]{Moezedin Javad Rafiee, MD}
\author[4]{Anastasia Oikonomou, MD}
\author[5]{Faranak Babaki Fard, MD}
\author[6]{Kaveh Samimi, MD}
\author[7]{Konstantinos N. Plataniotis}
\author[1,*]{Arash Mohammadi}

\affil[1]{Concordia Institute for Information Systems Engineering (CIISE), Concordia University, Montreal, Canada}
\affil[2]{Department of Electrical and Computer Engineering, Concordia University, Montreal, QC, Canada}
\affil[3]{Department of Medicine and Diagnostic Radiology, McGill University Health Center-Research Institute, Montreal, QC, Canada}
\affil[4]{Department of Medical Imaging, Sunnybrook Health Sciences Centre, University of Toronto, Toronto, Canada}
\affil[5]{Faculty of Medicine, University of Montreal, Montreal, QC, Canada}
\affil[6]{Department of Radiology, Iran university of medical science, Tehran, Iran}
\affil[7]{Department of Electrical and Computer Engineering, University of Toronto, Toronto, Canada}

\affil[*]{corresponding author: Arash Mohammadi (arash.mohammadi@concordia.ca)}

\begin{abstract}
Novel Coronavirus (COVID-19) has drastically overwhelmed more than 200 countries affecting millions and claiming almost 1 million lives, since its emergence in late 2019. This highly contagious disease can easily spread, and if not controlled in a timely fashion, can rapidly incapacitate healthcare systems. The current standard diagnosis method, the Reverse Transcription Polymerase Chain Reaction (RT- PCR), is time consuming, and subject to low sensitivity. Chest Radiograph (CXR), the first imaging modality to be used, is readily available and gives immediate results. However, it has notoriously lower sensitivity than Computed Tomography (CT), which can be used efficiently to complement other diagnostic methods. This paper introduces a new COVID-19 CT scan dataset, referred to as COVID-CT-MD, consisting of not only COVID-19 cases, but also healthy and subjects infected by Community Acquired Pneumonia (CAP). COVID-CT-MD dataset, which is accompanied with lobe-level, slice-level and patient-level labels, has the potential to facilitate the COVID-19 research, in particular COVID-CT-MD can assist in development of advanced Machine Learning (ML) and Deep Neural Network (DNN) based solutions.
\end{abstract}
\begin{document}

\flushbottom
\maketitle

\thispagestyle{empty}

\section*{Background \& Summary}
Since its first emergence in late 2019, novel Coronavirus (COVID-19) has drastically changed the world, impacting several aspects of the modern life. According to the World Health Organization (WHO), as of September 2020, more than 200 countries have confirmed positive COVID-19 cases, leading to more than 30 million cases and almost 1 million reported fatalities.  Considering statistics and impacts together with the fact that COVID-19 can easily spread if infected cases are not isolated/treated in a timely fashion, sensitive and accessible diagnosis systems are of significant importance. Reverse Transcription Polymerase Chain Reaction (RT- PCR)~\cite{XU:2020,Wang:2020}, which is currently considered as the gold standard diagnosis technique, suffers from relatively low sensitivity~\cite{Ai:2020}. Furthermore, It was not easily accessible in the beginning of epidemic in  most of the countries . More importantly, this test is time consuming that is not desirable as time is a critical factor in isolating, treating, and preventing the transition of COVID-19. Being able to identify COVID-19-related respiratory complications, Chest Radiographs (CXR)~\cite{Cozzi:2020}, can play an important complementary role for the RT- PCR test to asses complications. Although, CXR can act as a quantitative method to assess the extent of COVID-19 involvement and estimate the risk of Intensive Care Unit (ICU) admission, it still has lower sensitivity compared to Computed Tomography (CT)~\cite{Borakati:2020}. Due to high sensitivity and rapid access, chest CT plays a significant role in diagnosis and management of COVID-19 and has been recognized as the most sensitive imaging modality to detect complications.

Despite the high potential of CT  in contributing to the COVID-19 research and clinical usage, publicly available datasets are mostly limited to a few number of cases, are not accompanied with other types of respiratory diseases to facilitate comparisons, and are not associated with suitable labels. Furthermore, cases may be collected from different sources with different imaging protocols, limiting a unified study. In a few identified datasets, available CT scans are limited to only infected slices, rather than the complete volume. Another important aspect that should be considered in the available datasets is that whether labels are available in a patient-level, slice-level, and lobe-level fashion. The later can further contribute to identify the location of the COVID-19 infection. Finally, different types of labels and information, suitable for different tasks, are provided in identified datasets. Table~\ref{tab:datasets} provides an overview of the available datasets along with the provided COVID-19 related information.

The introduced COVID-19 CT scan dataset, referred to as the $\CTMD$, is applicable in Machine Learning (ML) and deep learning studies of COVID-19 classification. In particular, $\CTMD$ dataset consists of 171 confirmed positive COVID-19 cases (gathered from 2020/02/23 to 2020/04/21), 76 normal cases (gathered from 2019/01/21 to 2020/05/29), and 60 Community Acquired Pneumonia (CAP) cases (gathered from 2018/04/03 to 2019/11/24). All these cases are collected from Babak Imaging Center in Tehran, Iran, and labeled by three experienced radiologists in patient-level, slice-level, and lobe-level manners. Patient-level label refers to a single diagnosis assigned to the subject, whereas slice-level and lobe-level refer to identifying slices and lobes demonstrating infection, respectively. More importantly, the whole CT volume is available for all the subjects. $\CTMD$ is presented in Table~\ref{tab:datasets}, along with the previous datasets, to highlight its differences. Regarding Reference~\cite{Rahimzadeh:2020}, we would like to mention that while this Reference provides only COVID-19 and normal cases,  $\CTMD$ provides CAP cases additionally. Furthermore, $\CTMD$  is the only classification-related dataset that contains lobe-level information, which can significantly improve and contribute to the localization and analysis of  the COVID-19 infection.

\section*{Methods}
This section provides a description of the data collection procedure, inclusion criteria, and de-identification. Furthermore, detailed statistics of the data is presented to facilitate its usage. More importantly, applicability of the $\CTMD$ dataset for development of ML/DNN solutions is explained. This section is concluded by describing the possible limitations of the provided dataset. This research work is performed based on the policy certification number $30013394$ of Ethical acceptability for secondary use of medical data approved by Concordia University, Montreal, Canada. Furthermore, informed consent is obtained from all the patients.

\subsection*{Data Collection}
The $\CTMD$ dataset contains volumetric chest CT scans of 171 patients positive for COVID-19 infection, 60 patients with CAP, and 76 normal patients. COVID-19 cases are collected from February 2020 to April 2020, whereas CAP cases and normal cases are collected from April 2018 to December 2019 and January 2019 to May 2020, respectively, in Babak Imaging Center, Tehran, Iran. Diagnosis of COVID-19 infection is based on positive real-time Reverse Transcription Polymerase Chain Reaction (rRT-PCR) test results, clinical parameters, and CT scan manifestations identified by three thoracic radiologists, with more than 20, 18, and 25 years of experience in thoracic imaging, respectively. Labels provided by the three radiologists showed high degree of agreement (more than 90\%). Diagnosis for CAP and normal cases was confirmed using clinical laboratory tests, and CT scans. A subset of 55 COVID-19, and 25 CAP cases were analyzed by the radiologists to identify and label slices with evidence of infection. The labeled subset of the data contains 4,993 number of slices demonstrating infection and 18,416 number of slices without infection.

\begin{figure}[ht]
\centering
\includegraphics[scale=0.6]{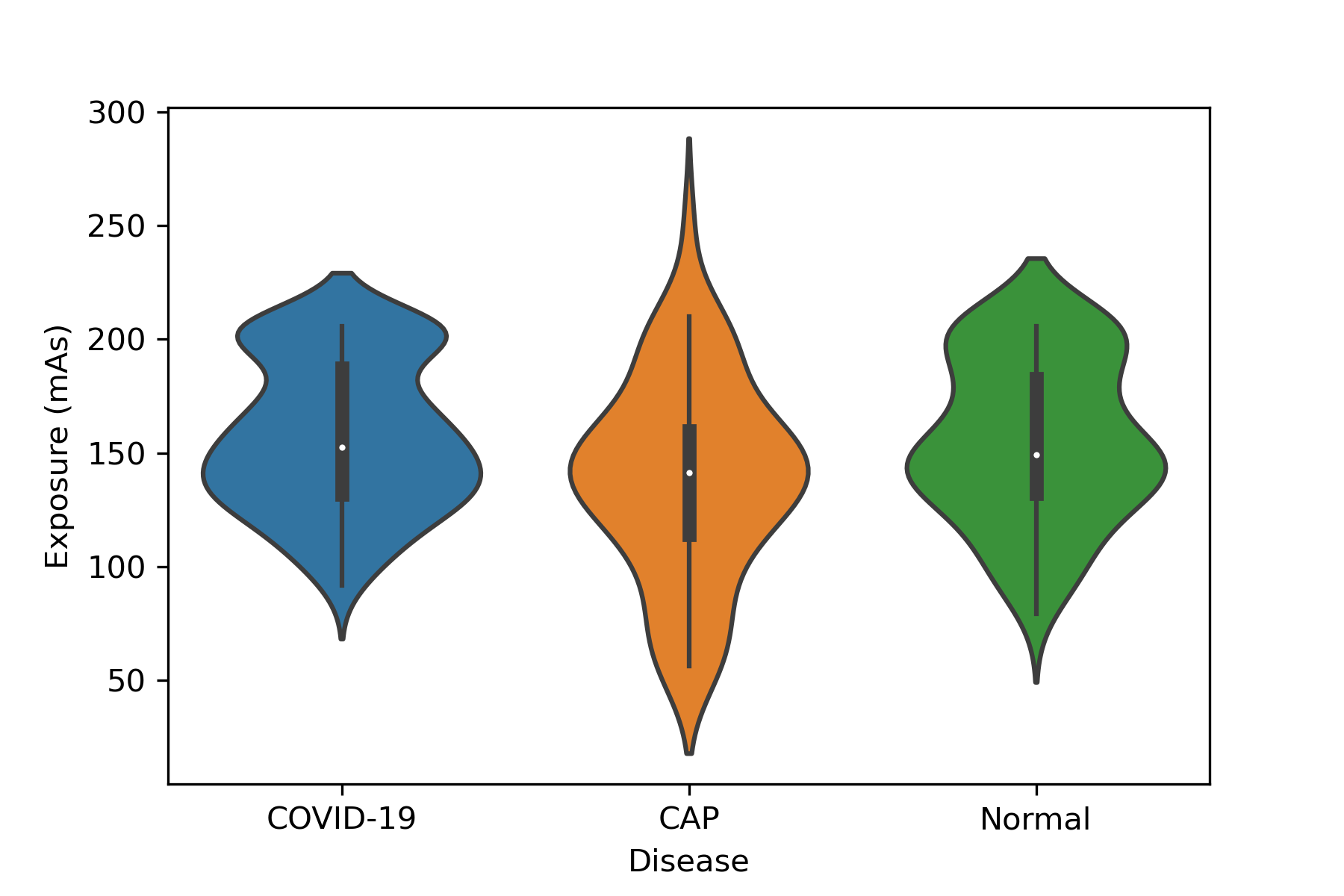}
\caption{The distribution of the Exposure values for COVID-19, CAP and Normal cases.}
\label{fig: exp}
\end{figure}

All images are axial, with a reconstruction matrix (output size of the images) of $512 \times 512$, and are obtained from a SIEMENS, SOMATOM Scope scanner. Table~\ref{tab:ct} shows  different CT acquisition settings, where Peak KiloVoltage  (kVp) and Exposure Time affect the radiation exposure dose, while slice thickness represents the axial resolution. As shown in Table~\ref{tab:ct}, slice thickness, kVP, and exposure time are almost the same with a few variations in a few CAP cases. Distance of Source to detector and Distance of Source to patient, which are traditionally referred to as SID and SOD, respectively, are also the same in all cases except for a few CAP cases. The minimum and maximum exposure value (in mAs) used in the scanning process is also presented in Table~\ref{tab:ct}. The exposure value determines the total radiation dose in CT scan. The distribution of the exposure values is illustrated by the violin plots for each disease type in Figure~\ref{fig: exp}. Accordingly, the mean and standard deviation of the exposure values are reported in Table~\ref{tab:exposure}.

\subsubsection*{CT Acquisition Care in The Medical Imaging Department}
As COVID-19 is highly contagious, all the staff of the medical imaging department involved in the CT acquisition  are provided with personnel protective equipment (PPE). More importantly, there is a minimum of 5-minute time slack between two consecutive CT scans, allowing enough time to sanitize the CT scanner.

\subsection*{Data Inclusion and Exclusion Criteria}
All cases with confirmed clinical diagnosis are included in the dataset. Nevertheless, during the data collection procedure, there were some cases related to the late 2019, with manifestations similar to those of COVID-19. However, as the first COVID-19 case in Iran is reported in early February 2020, these cases were excluded from the dataset. Furthermore, according to the radiologists' assessment, images with poor quality and visible artifacts were excluded.

\subsection*{De-identification}
To respect the patients' privacy, we de-identified all the CT studies  by removing the patients' name, birthday, and the name and address of the imaging center, as well as the operators' name from their headers. Some helpful information including patients' gender and age, the scanner type, and the image acquisition settings has been kept to preserve the statistical characteristics of the dataset.

\subsection*{Data Statistics}
\begin{figure}[ht]
\centering
\mbox{\subfigure[]{\includegraphics[scale=0.6]{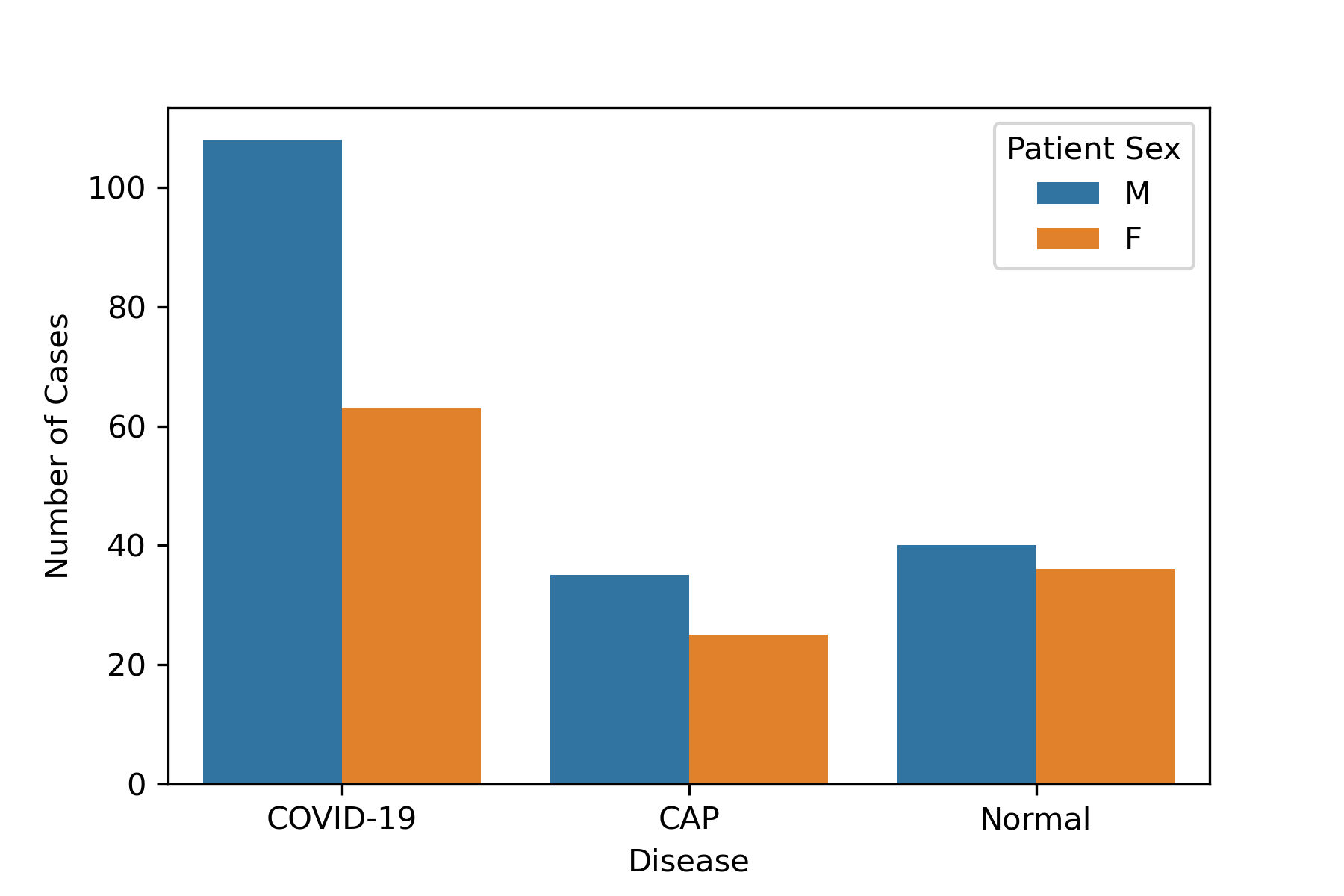}}
\subfigure[]{\includegraphics[scale=0.6]{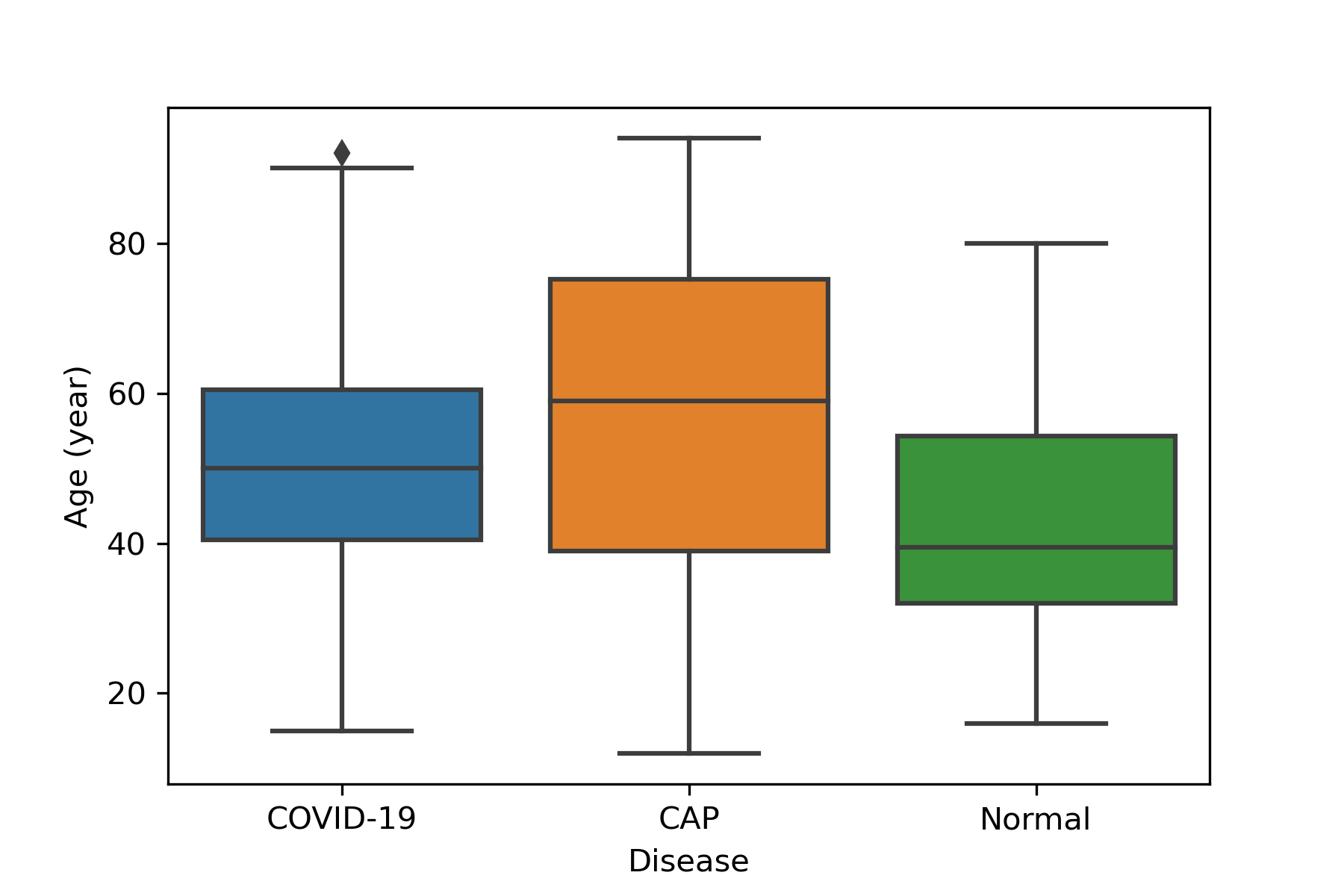}}}
\caption{\small (a) The number of cases separated by the patient's gender. (b) The distribution of age for COVID-19, CAP and Normal cases.}
\label{fig: sex}\label{fig: age}
\end{figure}

The demographic distribution of the dataset describing the gender and age distributions is illustrated in Table~\ref{tab:age} and Figure~\ref{fig: sex}. As shown in Figure~\ref{fig: sex}(a), males outnumbered females in this dataset. The boxplot in Figure~\ref{fig: age}(b) represents the important statistical parameters of the patients' age distribution. As shown in this boxplot, normal cases are mainly distributed in lower ages, while CAP cases are distributed in a wide range of ages with a higher average age.

\begin{figure}[ht]
\centering
\mbox{\subfigure[]{\includegraphics[scale=0.6]{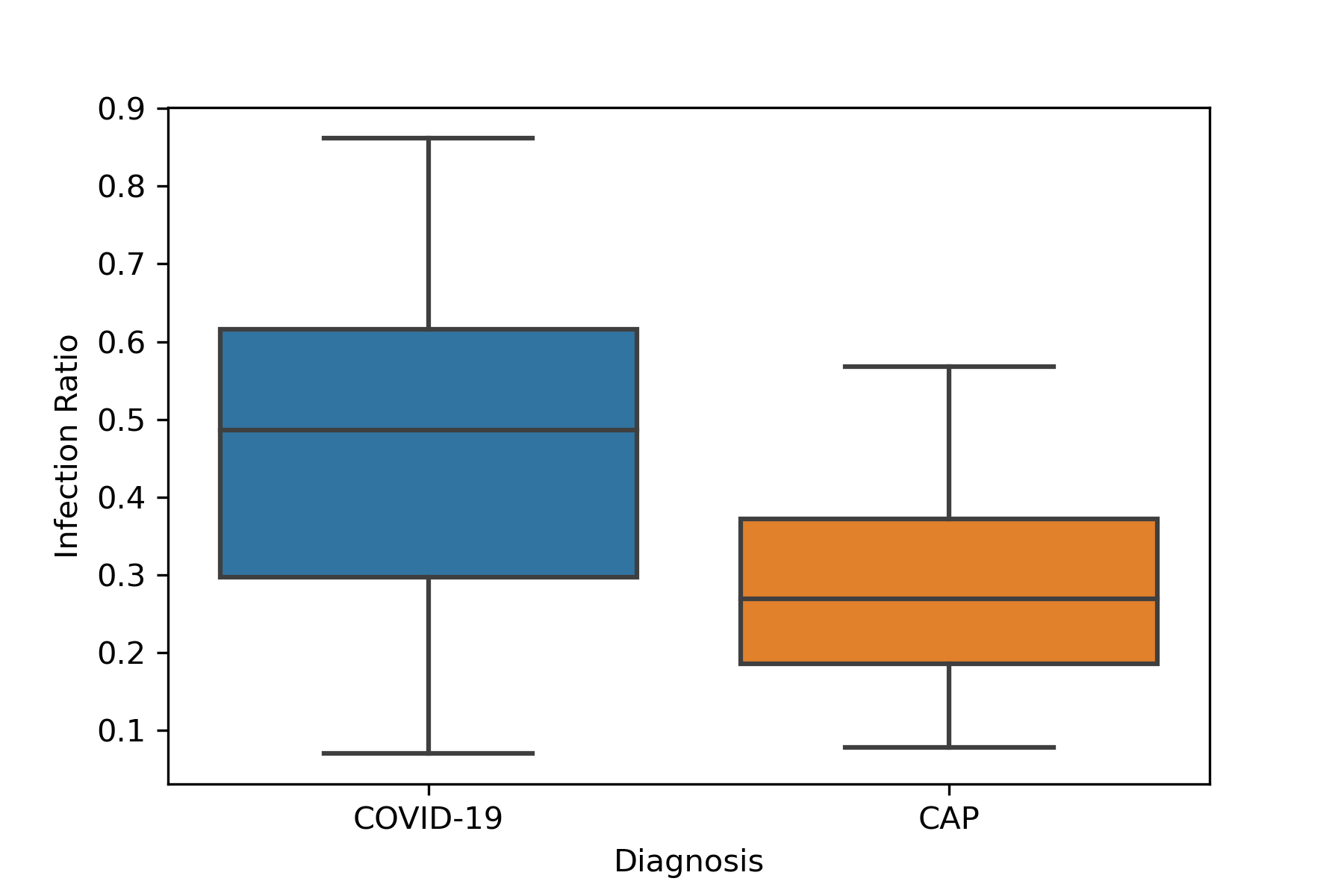}}
\subfigure[]{\includegraphics[scale=0.6]{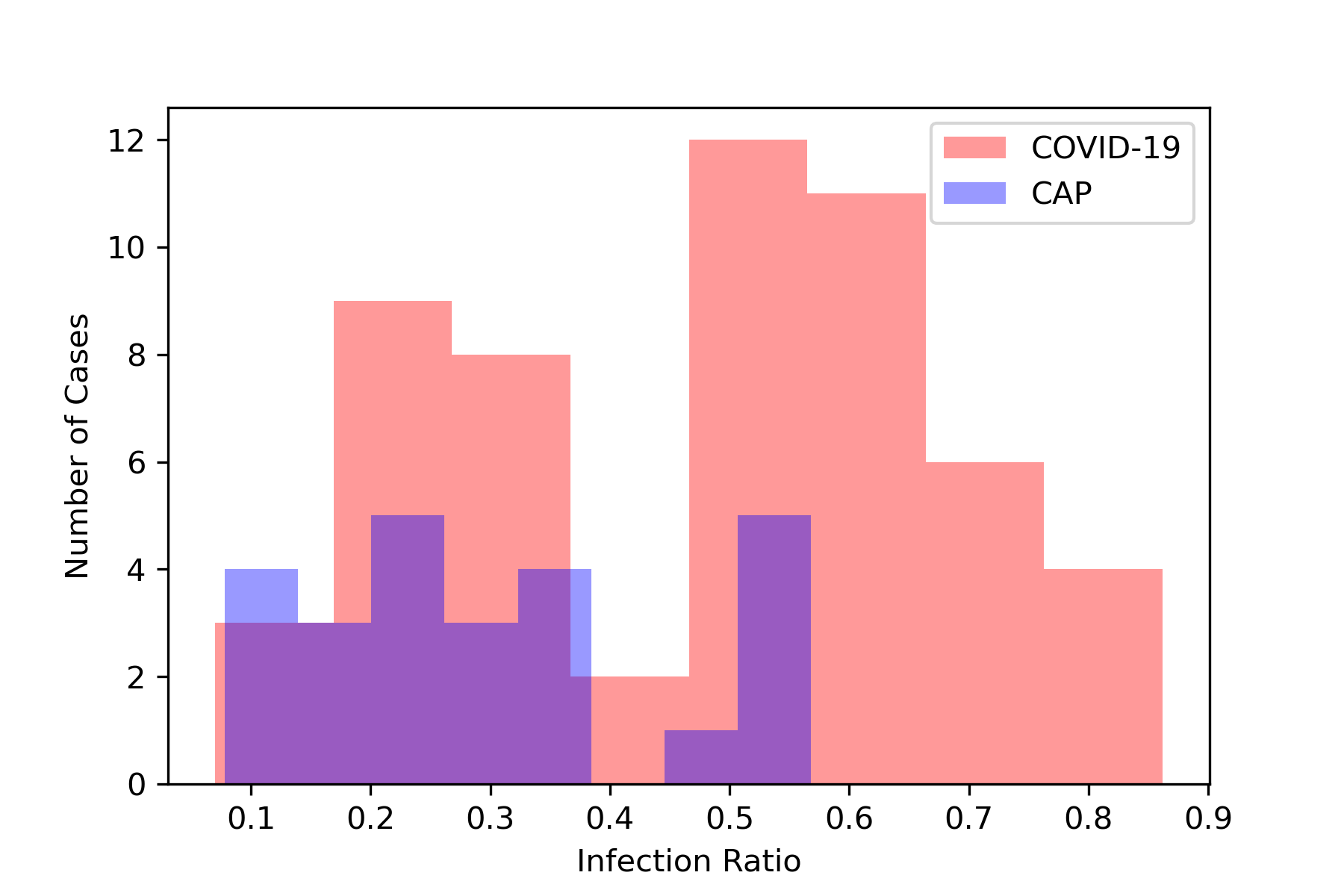}}}
\caption{\small (a) The distribution of the Infection Ratio in the labeled dataset for COVID-19 and CAP cases. (b) The histogram of the Infection Ratio in the labeled dataset for COVID-19 and CAP cases.}
\label{fig: rate}\label{fig: hist}
\end{figure}

As previously stated, part of the dataset is analyzed and the slice-level labels are extracted. The number of labeled cases and slices demonstrating infection are presented in Table~\ref{tab:infection}. Infection ratio in this table represents the ratio of the slices demonstrating infection to the total number of slices in a CT scan, which varies for different cases based on the severity and stage of the disease. The minimum and maximum values for the infection ratio in the labeled dataset are presented in Table~\ref{tab:infection}. The distribution of the Infection Ratio is also illustrated by the boxplots in Figure~\ref{fig: rate}(a), which demonstrate a higher infection ratio in COVID-19 cases compared to CAP cases. The histogram of the Infection Ratio values is illustrated in Figure~\ref{fig: hist}(b).

\begin{figure}[ht]
\centering
\includegraphics[scale=0.6]{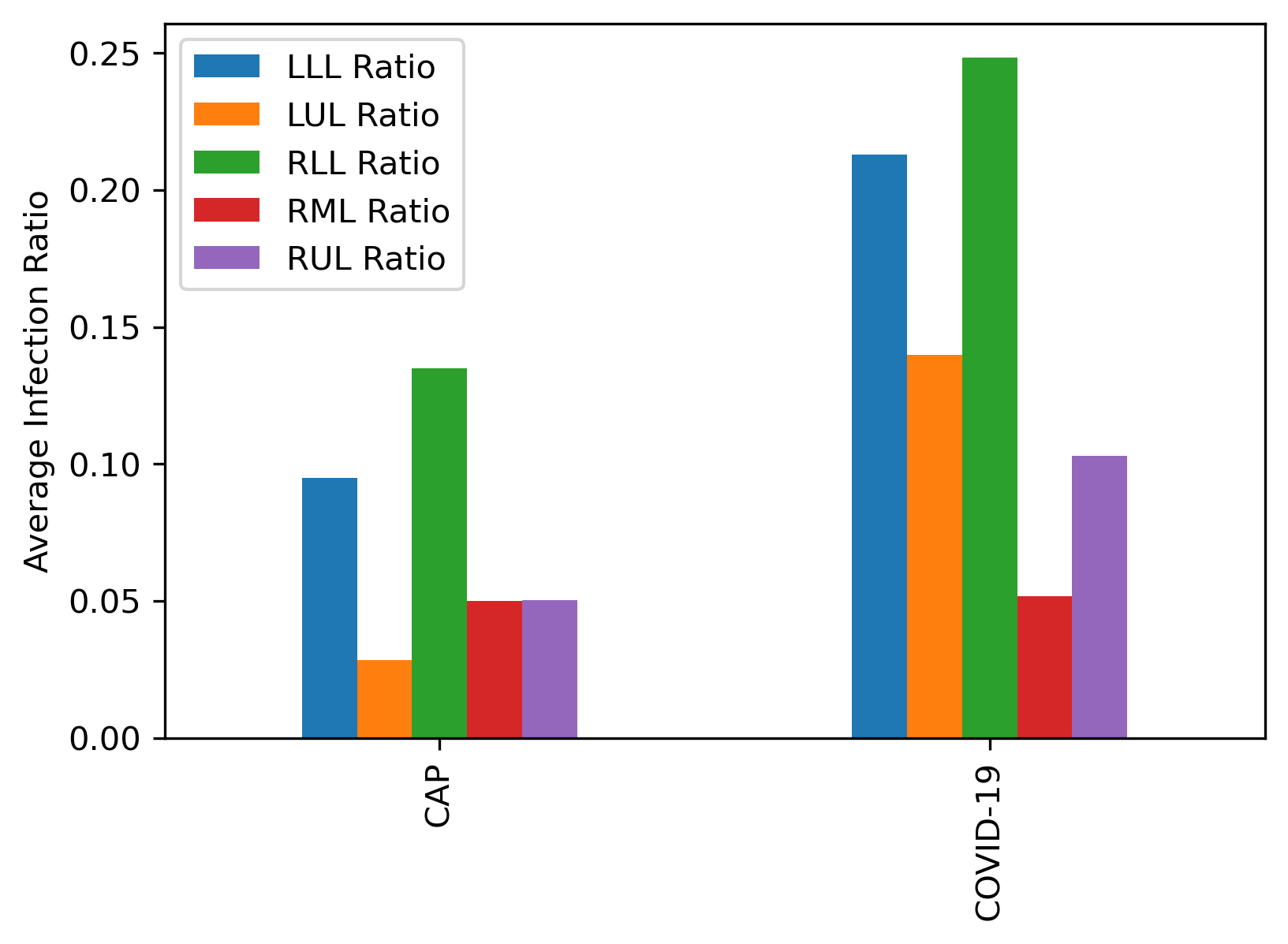}
\caption{Average Infection Ratio in each lobe of the lung for COVID-19 and CAP cases in the labeled dataset.}
\label{fig:rate_lobe}
\end{figure}

In addition to the described slice-level labels, the detailed distribution of infection in each lobe of the lung is provided by the radiologists. Table~\ref{tab:lobe} indicates the number of cases and slices with infection demonstrated in specific lung regions. Similar to Figure~\ref{fig: rate}, where the infection ratio was presented for the total slices with infection in the lung, the average of lobe infection ratios are presented in Figure~\ref{fig:rate_lobe}, illustrating the average ratio of slices demonstrating infection in a particular lobe to the total number of slices in a CT scan. As evident in Table~\ref{tab:lobe} and Figure~\ref{fig:rate_lobe}, the average infection ratio in the lower lobes is higher in both COVID-19 and CAP cases compared to other lung regions in our labeled dataset.

\subsection*{Limitations}
Although all cases and labels are confirmed by three experienced radiologists, we would like to describe a few limitations that the data users may encounter. These limitations are as follows:
\begin{itemize}
\item The slice and lobe labeling processes focus more on regions with distinctive manifestations rather than minimal findings.
\item Not all the COVID-19 patients have confirmed positive RT-PCR result, as this test was not publicly accessible in Iran at the time of the first emergence of the COVID-19. Furthermore, the high load of patients in need of COVID-19 examination, did not allow for an inclusive RT-PCR test. The diagnosis of some patients in the COVID-CT-MD dataset is confirmed based on the CT findings, as well as the clinical results.
\item Although most of the cases with low quality CT scans are excluded, there may still be some cases with mild motion artifact which is inevitable, since COVID-19 patients suffer from dyspnea.
\item During the slice and lobe labeling process, some suspicious areas adjacent to the chest wall and diaphragm are not labeled as ``infected'', due to their poor distinction.
\end{itemize}

\section*{Data Records}

\begin{figure}[h!]
\begin{forest}
  for tree={
    font=\ttfamily,
    grow'=0,
    child anchor=west,
    parent anchor=south,
    anchor=west,
    calign=first,
    inner xsep=7pt,
    forked edges,
    edge path={
        \noexpand\path [draw, \forestoption{edge}]
        (!u.south west) +(7.5pt,0) |- (.child anchor) pic {folder} \forestoption{edge label};
    },
    before typesetting nodes={
        if n=1
        {insert before={[,phantom]}}
        {}
    },
    fit=band,
    before computing xy={l=15pt},
  }
[Main Folder
  [COVID-19 subjects
    [Subject-ID
      [Slice-ID.dcm
      ]
    ]
]
  [Cap subjects
    [subject-ID
      [Slice-ID.dcm
      ]
    ]
 ]
   [Normal subjects
    [subject-ID
      [Slice-ID.dcm
      ]
    ]
 ]
 [Index.csv
 ]
 [Slice-level-labels.npy
 ]
 [Lobe-level-labels.npy
 ]
]
\end{forest}
\caption{Structure of the data included in $\CTMD$ dataset. \label{fig:nDatabase}}
\end{figure}
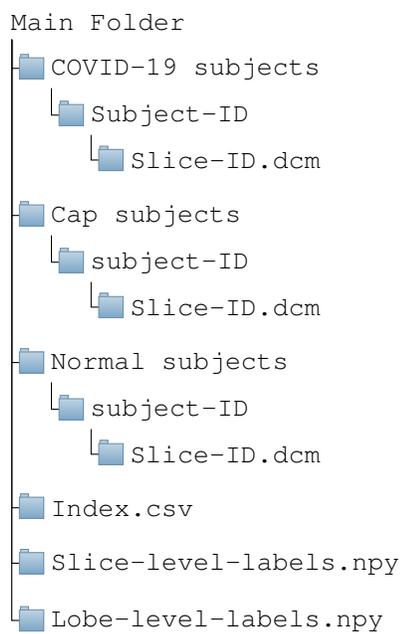

The diagram in Figure~\ref{fig:nDatabase} shows the structure of the $\CTMD$ dataset shared through Figshare\footnote{$\CTMD$ dataset is accessible through Figshare (\hyperlink{https://figshare.com/s/c20215f3d42c98f09ad0}{https://figshare.com/s/c20215f3d42c98f09ad0})}. COVID-19, CAP and Normal subjects are placed in separate folders, within which patients are arranged in folders, followed by CT scan slices in DICOM format. ``Index.csv'' is related to the patients having slice-level and lobe-level labels. The indices given to patients in ``Index.csv'' file are then used in ``Slice-level-labels.npy'' and ``Lobe-level-labels.npy'' to indicate the slice and lobe labels. ``Slice-level-labels.npy'' is a 2D binary Numpy array   in which the existence of infection in a specific slice is indicated by 1 and the lack of infection is shown by 0. In ``Slice-level-labels.npy'', the first dimension represents the case index and the second one represents the slice numbers. ``Lobe-level-labels.npy'' is a 3D binary Numpy array in which the existence of infection in a specific lobe and slice is determined by 1 in the corresponding element of the array. Like the slice-level array, in ``Lobe-level-labels.npy'', the two first dimensions represent the case index and slice numbers respectively. The third dimension shows the lobe indices which are specified as follows:
\begin{itemize}
\item 0 : Left Lower Lobe (LLL)
\item 1 : Left Upper Lobe (LUL)
\item 2 : Right Lower Lobe (RLL)
\item 3 : Right Middle Lobe (RML)
\item 4 : Right Upper Lobe (RUL)
\end{itemize}

\section*{Technical Validation}
Two noteworthy parameters in the studies using CT scans are the quality control and calibration of the scanning device. The longest time period between the scanner auto-calibration and the study in the $\CTMD$ dataset is 1 day, which ensures calibrated and accurate performance of the scanning device. Furthermore, there is an annual SIEMENS quality control that ensures the absence of ring artifacts in the acquired CT scans.

\section*{Usage Notes}
With the increasing number of COVID-19 patients, healthcare workers are overwhelmed with a heavy workload, lowering their concentration for a proper diagnosis. Accurate and timely COVID-19 diagnosis, on the other hand, is a critical factor in preventing the disease transition, treatment, and resource allocation. Machine Learning (ML), in particular Deep Learning (DL) based on Deep Neural Networks (DNN), is shown to be practical and effective in COVID-19 diagnosis and severity assessment. The $\CTMD$ dataset is specifically designed to facilitate  application of ML/DL in COVID-19-related tasks. In particular, this dataset can be used towards:
\begin{itemize}
\item A patient-level binary classification~\cite{Ozturk:2020,Afshar:2020} to distinguish COVID-19 from all other cases.
\item A patient-level multi-class classification~\cite{Ozturk:2020} to identify COVID-19, CAP, and normal subjects.
\item A slice-level~\cite{Yan:2020} and lobe-level classification to separate infected slices and lobes from non-infected ones for further analysis.
\item Slice-level and lobe-level labels can be used as additional inputs to segmentation models~\cite{Fan:2020}, to focus on only infected slices.
\item Slice-level and lobe-level labels can be used in generative models to generate artificial COVID-19 images, towards increasing the security of the healthcare systems and developing attack resilient solutions~\cite{Mirsky:2020}.
\end{itemize}
We have utilized the $\CTMD$ dataset in a recent study~\cite{Heidarian:2020}, to classify subjects as COVID-19 or non-COVID (Normal and CAP). The model proposed in this study, referred to as the COVID-FACT, consists of two stages. In the first stage, infected slices (COVID-19 and CAP) are separated from healthy ones, through a developed Capsule Network. Consequently, in the second stage, infected slices are used to classify patients as COVID-19 or non-COVID, through another Capsule Network and an average voting approach. While the first stage exploits the provided slice-level labels, the patient-level ones are used in the second stage. Data users are encouraged to train and test their methods on the COVID-CT-MD dataset and compare their results, accordingly.

\section*{Code availability}
The Python code used to generate the statistical analysis and plots is shared within the same Figshare link\\ (\hyperlink{https://figshare.com/s/c20215f3d42c98f09ad0}{https://figshare.com/s/c20215f3d42c98f09ad0}).

\bibliography{NSD-vf}

\section*{Acknowledgements}
This work was partially supported by the Natural Sciences and Engineering Research Council (NSERC) of Canada through the NSERC Discovery Grant RGPIN-2016-04988.

\section*{Author contributions statement}
P.A. drafted the manuscript together with A.M., and analyzed the results. Sh.H. performed the statistical analysis and organized the dataset. M.J.R., F.B.B., and K.S. collected the dataset and revised the medical information in the paper.  N.E., F.N., A.O., K.N.P. edited and revised the manuscript. A.M. supervised the study.

\section*{Competing interests}
The authors declare no competing interests.

\section*{Figures \& Tables}

\begin{table}[ht]

\centering
\caption{\label{tab:datasets}Available COVID-19 CT scan datasets. NA stands for not available.}
\renewcommand{\arraystretch}{2}
\begin{adjustbox}{width=1\textwidth}
\begin{tabular}{c c|c|c|c|c|c|c|c|c|c|c|c|}

\cline{2-13}

 & \multicolumn{3}{|c|}{\textbf{Number of cases}}
 & \multicolumn{2}{|c|}{\textbf{Label type}}
 & \multicolumn{2}{|c|}{\textbf{Data Source}}
 & \multicolumn{2}{|c|}{\textbf{CT volume}}
  & \multicolumn{3}{|c|}{\textbf{Label Level }}
 \\
\hline
 \multicolumn{1}{|c|}{Dataset} & COVID & CAP & Normal  & Classification & Segmentation  & Multiple & Single  & Available & Not available & Patient-level & Slice-level & Lobe-level \\
\hline
\multicolumn{1}{|c|}{Reference~\cite{Bjorke:2020}} & 49 & NA & NA & &\cmark  & \cmark & & \cmark & & &\cmark & \\
\hline
\multicolumn{1}{|c|}{Reference~\cite{Jun:2020}} & 20 & NA & NA & &\cmark  & \cmark & & \cmark & & &\cmark & \\
\hline
\multicolumn{1}{|c|}{Reference~\cite{Cohen:2020}} & 20 & NA & NA & &\cmark  & \cmark & & \cmark & & &\cmark & \\
\hline
\multicolumn{1}{|c|}{Reference~\cite{Morozov:2020}} & 856 & NA & 254 & \cmark &  & \cmark & & \cmark & & \cmark& & \\
\hline
\multicolumn{1}{|c|}{Reference~\cite{Zhao:2020}} & 216 & NA & 55 & \cmark &  & \cmark & &  &\cmark & &\cmark & \\
\hline
\multicolumn{1}{|c|}{Reference~\cite{Soares:2020}} & 60 & NA & 60 & \cmark &  & \cmark & &  &\cmark & &\cmark & \\
\hline
\multicolumn{1}{|c|}{Reference~\cite{Rahimzadeh:2020}} & 95 & NA & 282 & \cmark &  &  &\cmark & \cmark & & \cmark&\cmark & \\
\hline
\multicolumn{1}{|c|}{COVID-CT-MD} & 171 & 60 & 76 & \cmark &  &  &\cmark & \cmark & & \cmark&\cmark & \cmark\\
\hline
\end{tabular}
\end{adjustbox}
\end{table}

\begin{table}[ht]
\centering
\caption{CT scan settings used to acquire the COVID-CT-MD dataset.}
\label{tab:ct}
\renewcommand{\arraystretch}{2}
\begin{adjustbox}{width=1\textwidth}
\begin{tabular}{|c|c|c|c|c|c|c|c|}
\hline
Diagnosis  & Slice Thickness (mm) & Peak Kilovoltage (kVp)  & Exposure Time (ms) & X-ray Tube Current (mA) & SID (mm) & SOD (mm) & Exposure values (mAs)\\
\hline
\textbf{COVID-19}   & 2 & $110-130$ & 600 & $153-343$ & 940 & 535 & $91.8-205.8$ \\
\hline
\textbf{CAP}  & 2 & $110-120$ & $420-600$ & $94-500$ & $940-1040$ & $535-570$ & $56.4-250.0$\\
\hline
\textbf{Normal}  & 2 & $110$ & 600 & $132-343$ & 940 & 535 & $79.2-205.8$\\
\hline
\end{tabular}
\end{adjustbox}
\end{table}

\begin{table}[ht]
\centering
\caption{The statistical parameters (mean and standard deviation) of the Exposure values.}
\label{tab:exposure}
\renewcommand{\arraystretch}{2}
\begin{tabular}{|c|c|c|}
\hline
Diagnosis & Exposure mean & Exposure standard deviation \\
\hline
\textbf{COVID-19} & $157.15$ & $32.64$ \\
\hline
\textbf{CAP}  & $138.96$ & $43.15$ \\
\hline
\textbf{Normal} & $155.12$ & $35.24$  \\
\hline
\end{tabular}
\end{table}

\begin{table}[ht]
\centering
\caption{Gender and age distribution in COVID-CT-MD}
\label{tab:age}
\renewcommand{\arraystretch}{2}
\begin{tabular}{|c|c|c|c|}
\hline
Diagnosis & Cases & Sex & Age (year) \\
\hline
\textbf{COVID-19} & 171 & 108 M/63 F & $51.6 \pm 14.6$ \\
\hline
\textbf{CAP} & 60 & 35 M/25 F & $57.7 \pm 21.7$ \\
\hline
\textbf{Normal} & 76 & 40 M/36 F & $43.4 \pm 14.1$ \\
\hline
\end{tabular}
\end{table}

\begin{table}[ht]
\centering
\caption{The number of cases, Slices, and Infection Ratio in the labeled dataset.}
\label{tab:infection}
\renewcommand{\arraystretch}{2}
\begin{tabular}{|c|c|c|c|c|}
\hline
Diagnosis & Cases & Slices Demonstrating Infection & Slice without infection & Infection Ratio \\
\hline
\textbf{COVID-19} & 55 & 3815 & 4377 & $7.0\%-86.2\%$ \\
\hline
\textbf{CAP}  & 25 & 1178 & 2718 & $7.8\%-56.8\%$ \\
\hline
\end{tabular}
\end{table}

\begin{table}[ht]
\centering
\caption{Number of cases and slices, respectively, demonstrating infection in each lobe. LLL: Left Lower Lobe – LUL: Left Upper Lobe – RLL: Right Lower Lobe and Lingula – RML: Right Middle Lobe – RUL: Right Upper Lobe}
\label{tab:lobe}
\renewcommand{\arraystretch}{2}
\begin{tabular}{|c|c|c|c|c|c|}
\hline
Diagnosis & LLL & LUL & RLL & RML & RUL \\
\hline
\textbf{COVID-19} & $43\&1705$ &	$38\& 1120$ &	$45\& 2008$	&  $26\& 420$	& $29\& 826$ \\
\hline
\textbf{CAP}  & $13\& 374$	 & $5\& 117$	& $18\& 519$	 & $7\& 186$	& $9\& 208$ \\
\hline
\textbf{Total}& $56\& 2079$	& $43\& 1237$	& $63\& 2527$& 	$33\& 606$	& $38\& 1034$\\
\hline
\end{tabular}
\end{table}

\end{document}